\documentclass[cameraready]{Interspeech}

\title{DASH: Dual-View Self-Distillation with Multi-Layer Hidden Representations for Robust Speech Recognition}


\usepackage{comment}
\usepackage{multirow}
\usepackage{tabularx}
\usepackage{color, colortbl}
\definecolor{Gray}{gray}{0.9}
\usepackage{subcaption}

\author[affiliation={1}, orcid=0009-0008-6531-6248, equalcontribution]{Jaeeun}{Baik$^\dagger$}
\author[affiliation={2}, orcid=0000-0002-6145-7157, equalcontribution]{Ui-Hyeop}{Shin}
\author[affiliation={1}]{Jiwon}{Lee}
\author[affiliation={1}, orcid=0009-0008-4094-8492]{Woocheol}{Jeong}
\author[affiliation={1,2}, orcid=0000-0002-7105-5493]{Hyung-Min}{Park}

\address{
    $^1$Department of Artificial Intelligence, Sogang University, Republic of Korea \\
    $^2$Department of Electronic Engineering, Sogang University, Republic of Korea
}

\email{\{jaeeunbaik, dmlguq123, jiwonnn, charmingwc, hpark\}@sogang.ac.kr}

\begin{document}

\maketitle

\begin{abstract}
Automatic Speech Recognition (ASR) often degrades in real-world noisy environments, making noise robustness essential for deployment. Supervised noise-augmented fine-tuning is a common remedy, but it can introduce a robustness--clean trade-off and overfit to specific corruptions, degrading recognition in clean conditions. We propose DASH, a self-distillation framework that improves robustness by learning clean--noisy consistency from paired views. DASH distills hidden representations from multiple encoder layers to capture features from low-level acoustics to high-level semantics, and stabilizes training by minimizing KL divergence between prototype assignment distributions of clean and noisy views. Experiments on LibriSpeech show that DASH consistently improves recognition under diverse noisy conditions while preserving clean accuracy, achieved by a label-free pre-training stage with minimal additional overhead (about 4\% of fine-tuning time) beyond standard fine-tuning.
    
\end{abstract}

\section{Introduction}
\begingroup
\renewcommand{\thefootnote}{}
\footnotetext{\hspace{-1.25mm}$^\dagger$Currently with Cochl.}
\addtocounter{footnote}{-1}
\endgroup
Recent advancements in automatic speech recognition (ASR) have been largely driven by the scaling up of model architectures and the utilization of massive training datasets, leading to significant performance improvements~\cite{radford_robust_2023, rekesh_fast_2023, liu_voxtral_2025, shi_qwen3-asr_2026}. Concurrently, the evolution of speech foundation models has further propelled ASR capabilities by leveraging self-supervised learning (SSL) to extract rich acoustic representations without the need for manual labels~\cite{liu_mockingjay_2020, baevski_wav2vec_2020, hsu_hubert_2021, baevski_data2vec_2022, chen_wavlm_2022}. Despite these remarkable achievements, ASR systems continue to suffer from severe performance degradation in noisy conditions, which remains a critical bottleneck for their reliable deployment in real-world applications. Moreover, naively incorporating noise into the supervised training process to adapt the model to noisy conditions often compromises its performance in clean environments.

Therefore, to enhance robustness against noise while preserving baseline accuracy in clean conditions, it is essential for ASR models to learn consistent feature representations across clean and noisy domains.
To achieve such feature consistency, self-distillation-based learning techniques, such as BYOL, have been explored~\cite{grill_bootstrap_2020}. Specifically, BYOL introduced a bootstrap-style, non-contrastive framework in which differently augmented views are processed by paired networks (one of which is a momentum target) to learn representations without relying on negative samples. Building upon this paradigm, subsequent methods like DINO and iBOT have achieved remarkable success in the computer vision domain~\cite{zhou_ibot_2021, caron_emerging_2021}.

Within the speech recognition domain itself, recent studies have emerged with a similar problem awareness, aiming to enhance robustness through consistency and contrastive learning paradigms. For instance, Speech-SimCLR~\cite{jiang_speech_2021} explored self-supervised representation learning by maximizing the agreement between differently augmented speech samples in the latent space, coupled with a reconstruction objective. Similarly, CR-CTC~\cite{yao_cr-ctc_2025} introduced consistency regularization directly into the Connectionist Temporal Classification (CTC) framework, enforcing the model to yield consistent output distributions across diverse augmented views of the input, which effectively mitigated overfitting to specific noise patterns. Furthermore, HuBERT-VIC~\cite{ahn_hubert-vic_2025} extended this concept to speech foundation models by employing Variance-Invariance-Covariance (VIC) regularization, explicitly aligning the representations of clean and noisy speech to ensure robust generalization across various noise conditions.

Building on these insights, we present DASH (\textbf{D}u\textbf{A}l-view \textbf{S}elf-distillation with multi-layer \textbf{H}idden representations), a framework that learns noise-invariant speech representations by distilling clean-to-noisy consistency across multiple encoder layers.
To achieve this, we employ a self-distillation approach with an Exponential Moving Average (EMA) to establish a stable teacher-student dynamic~\cite{tarvainen_mean_2017}. Furthermore, to circumvent gradient interference and delayed convergence in joint optimization, we adopt a decoupled two-stage training paradigm. Unlike hybrid joint-training approaches, the decoupled architecture allows the initial pre-training stage to operate without text labels, enabling DASH to leverage unlabeled speech data. The model first stabilizes its parameters through distillation-only pre-training, followed by supervised ASR fine-tuning for target task adaptation. DASH introduces only a lightweight encoder-only distillation stage (5k steps) before standard fine-tuning, adding a small computational overhead while improving noise robustness.
Finally, to ensure training stability and prevent trivial shortcuts, DASH optimizes prototype-based probability distributions via Kullback-Leibler (KL) divergence across multiple intermediate encoder layers, effectively capturing both low-level acoustics and high-level semantics \cite{chang_distilhubert_2022}.
Experiments on LibriSpeech~\cite{panayotov_librispeech_2015} show that DASH consistently enhances recognition under noisy conditions, indicating better generalization beyond fine-tuning alone.

\begin{figure*}[t]
  \centering
  \vspace{-3mm}
  \includegraphics[width=0.82\linewidth]{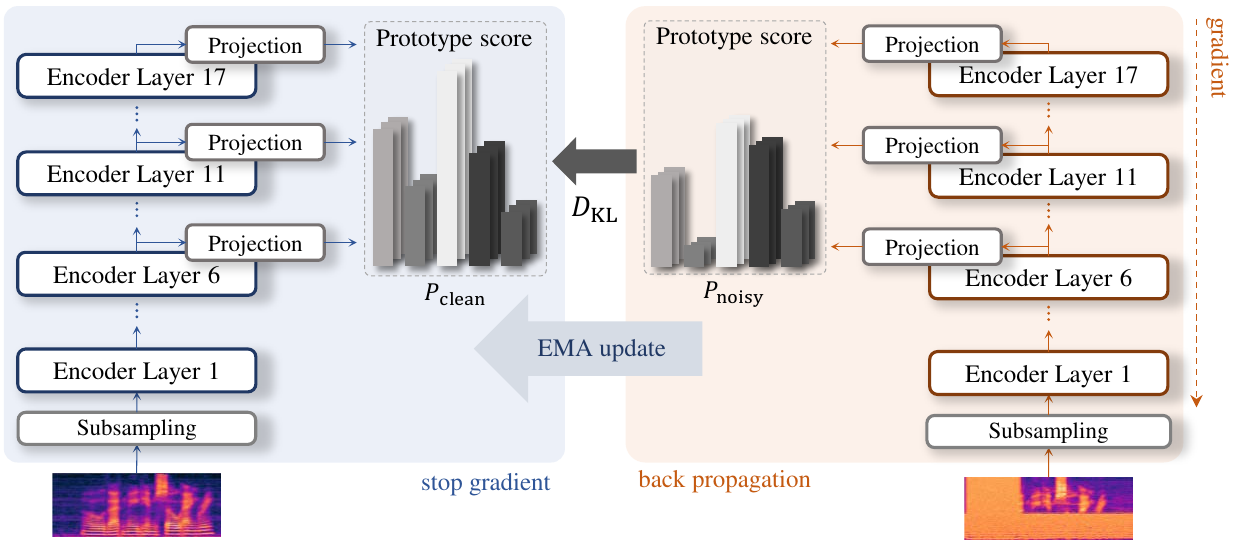}
  \vspace{-3mm}
  \caption{Overview of the DASH pipeline.}
  \vspace{-3mm}
  \label{fig:pipeline}
\end{figure*}

\section{DASH}

\subsection{Clean-Noisy Pair with Dual Encoder Branches}

Figure~\ref{fig:pipeline} illustrates the overall architecture of DASH. 
The core intuition behind DASH is to encourage the model to learn noise-invariant speech representations by comparing paired clean and noisy views of the same utterance. 
To achieve this, we instantiate a dual-branch encoder architecture consisting of a teacher (clean) network and a student (noisy) network. The clean network processes the unperturbed view $x_{\text{clean}}$, extracting clean acoustic features, while the noisy network processes the augmented, noisy view $x_{\text{noisy}}$. 

In this teacher-student framework, $x_{\text{clean}}$ serves as the input to the teacher network, providing stable and informative target representations. To ensure consistent and patient distillation \cite{beyer_knowledge_2022}, the weights of the clean network ($\theta_{\text{teacher}}$) are not updated via standard backpropagation. Instead, they are updated using an EMA of the student network's weights ($\theta_{\text{student}}$). This prevents the teacher's weights from fluctuating rapidly with each step-wise loss calculation, preserving historical parameter information while smoothly reflecting newly acquired knowledge:
\vspace{-1mm}
\begin{align}
    \theta_{\text{teacher}}^{(t+1)} = \alpha \cdot \theta_{\text{teacher}}^{(t)} + (1-\alpha) \cdot \theta_{\text{student}}^{(t)},
\end{align}
where $\alpha$ denotes the EMA decay rate. Because the clean network does not require gradient computation, we apply a stop-gradient operator to its outputs. By minimizing the divergence between the clean and noisy outputs, the student network is explicitly guided to focus on robust speech characteristics that are invariant to environmental noise. 
Formally, this process enables the baseline model to learn an invariant latent representation mapping $f : \mathcal{X} \to \mathcal{Y}$ that remains stable across different noise transformations $t \in \mathcal{T}$ applied to the input space $\mathcal{X}$, such that
\begin{align}
    f(t(x)) \approx f(x) \quad \text{for any } x \in \mathcal{X} \text{ and } t \in \mathcal{T}.
\end{align}

\subsection{Distillation Based on Prototypes}

A critical challenge when applying self-supervised learning objectives to predictive models in a supervised setting is the \textit{shortcut} problem, where the model bypasses learning meaningful representations in favor of trivial solutions, potentially degrading its primary objective of speech recognition \cite{geirhos_shortcut_2020, jiang_speech_2021}.
Simply minimizing the distance between continuous teacher and student logits often exacerbates this issue. This phenomenon is particularly problematic in ASR, where the model may exploit low-level acoustic correlations or specific noise patterns instead of capturing robust phonetic information.

To prevent representational collapse in the continuous feature space and avoid these superficial shortcuts, DASH introduces a prototype-based distillation approach. We employ a projection head followed by k-means clustering to quantize the continuous encoder outputs into discrete acoustic units. This transforms the continuous representation space into a structured categorical space. These discrete prototypes serve as robust targets for the self-supervised objective, effectively replacing continuous soft predictions and guiding the model to learn higher-level semantic structures.

%
%
\subsection{Training Objective}
As outlined in our two-stage training paradigm, the pre-training phase relies exclusively on the distillation objective. Let $f_{\theta}$ denote the encoder and $h_{\theta}$ the projection head. For an input $\mathbf{x} \in \mathbb{R}^{T \times D_{\text{in}}}$, the encoder produces hidden representations $f_{\theta}(\mathbf{x}) \in \mathbb{R}^{T \times D_{\text{enc}}}$.
To comprehensively capture both low-level acoustics and high-level semantics, we extract these hidden representations from multiple intermediate layers of the encoder (e.g., layers 6, 11, and 17, as shown in Figure~\ref{fig:pipeline}). These multi-layer outputs are independently passed through the projection head to obtain lower-dimensional embeddings:
\begin{equation}
    \mathbf{z} = h_{\theta}(f_{\theta}(\mathbf{x})) \in \mathbb{R}^{T \times D_{K}}.
\end{equation}

Next, we compute the similarity between each frame embedding $\mathbf{z}_{b,t} \in \mathbb{R}^{D_{K}}$ and a set of $K$ prototypes $\mathbf{C} \in \mathbb{R}^{K \times D_K}$, which are obtained via k-means clustering over a buffer of projection vectors. The frame-prototype similarity scores are then normalized into probability distributions using a softmax function with temperature $\tau_{temp}$: 
\vspace{-1mm}
\begin{equation}
    \begin{aligned}
        P_{\text{clean}}(k) &= \frac{\exp \bigl(\mathbf{z}^{\top}_{\text{clean}} \mathbf{c}_k / \tau_{temp} \bigr)}{\sum_{j=1}^{K} \exp \bigl(\mathbf{z}^{\top}_{\text{clean}} \mathbf{c}_j / \tau_{temp} \bigr) }, \\
        P_{\text{noisy}}(k) &= \frac{\exp \bigl(\mathbf{z}^{\top}_{\text{noisy}} \mathbf{c}_k / \tau_{temp} \bigr)}{\sum_{j=1}^{K} \exp \bigl(\mathbf{z}^{\top}_{\text{noisy}} \mathbf{c}_j / \tau_{temp} \bigr) }.
    \end{aligned}
\end{equation}

Finally, to enforce consistency across the dual views, we minimize the KL divergence from the stable clean distribution to the online noisy distribution. The overall DASH pre-training objective is formulated as
\begin{equation}\label{eqn:self-distillation}
    \mathcal{L}_{\text{DASH}} = \frac{1}{T} \sum_{t=1}^T D_{\text{KL}}\Bigl( \mathrm{sg}\bigl(P_{\text{clean}}\bigr) \parallel P_{\text{noisy}} \Bigr),
\end{equation}
where $\mathrm{sg}(\cdot)$ denotes the stop-gradient operator. This formulation ensures that gradients flow exclusively through the student (noisy) branch, encouraging it to produce prototype assignments that are invariant to acoustic perturbations while anchoring to the slowly evolving, clean targets of the teacher network.

\begin{table*}[!t]
\small
\vspace{-1mm}
\caption{WER (\%) of baseline, fine-tuning model with ASR(TDT-CTC) loss, and proposed DASH model on LibriSpeech test-clean/test-other and NOISEX-92 mixed conditions. All augmentation settings include SpecAug. When training in the phases, noise mixing is implemented in two ways, SNR range 0 to 15 dB or -5 to 10 dB. We implemented DASH on the baseline model which includes encoder-only training with unlabeled dataset (phase 1) and fine-tuning with labeled dataset (phase 2).}
\vspace{-3mm}
\renewcommand{\tabcolsep}{6pt}
\def\arraystretch{1.02}
\begin{center}

\scalebox{0.86}{\begin{tabular}{ccccccccccccc}
\toprule
    \multicolumn{2}{c}{\multirow{1}{*}{\textbf{Augmentation Method}}}
    & \multirow{2}{*}{\textit{test-clean}} & \multirow{2}{*}{\textit{test-other}}
    & \multicolumn{3}{c}{\normalsize\textit{test-clean + white}} 
    & \multicolumn{3}{c}{\normalsize\textit{test-clean + pink}}
    & \multicolumn{3}{c}{\normalsize\textit{test-clean + babble}}\\
\cmidrule(lr){1-2}\cmidrule(lr){5-7}\cmidrule(lr){8-10}\cmidrule(lr){11-13}
Phase 1 & Phase 2&&&0 dB & 5 dB & 10 dB & 0 dB & 5 dB & 10 dB & 0 dB & 5 dB & 10 dB\\
\midrule
\rowcolor{Gray}
\multicolumn{13}{c}{\textit{Baseline}}\\
- & - & 2.58 & 5.41
& 19.04 & 8.47 & 4.65 
& 19.79 & 7.11 & 4.01 
& 19.80 & 6.73 & 3.90 
\\
\midrule
\rowcolor{Gray}
\multicolumn{13}{c}{\textit{Fine-tuning only}}\\
- & Clean & 2.00 & 4.45 & 15.49 & 6.68 & 3.55 & 16.73 & 5.77 & 3.12 & 19.11 & 5.79 & 2.92\\
- & Noisy (0 to 15 dB) & 2.07 & 4.35
& 10.89 & 5.22 & 3.18
& 11.82 & 4.70 & 2.90
& 13.78 & 4.77 & 2.78
\\
- & Noisy (-5 to 10 dB) & 2.14 & 4.40
 & 10.34 & 5.02 & 3.14 
 & 11.12 & 4.50 & 2.83 
 & 13.06 & 4.50 & 2.71 
\\
\midrule
\rowcolor{Gray}
\multicolumn{13}{c}{\textit{DASH (self-distillation $\to$ fine-tuning})}\\
Noisy (0 to 15 dB) & Clean & 1.99 & \bf 4.10 & 
11.89 & 5.33 & 3.14 & 12.78 & 4.61 & 2.75 & 16.48 & 5.07 & 2.72
\\
Noisy (0 to 15 dB) & Noisy (0 to 15 dB) & \bf 1.96 & 4.15
 & \bf 10.27 & 4.88 & \bf 3.00 
 & 11.16 & \bf 4.40 & \bf 2.71 
 & 13.11 & 4.51 & \bf 2.64 
\\
Noisy (-5 to 10 dB) & Noisy (-5 to 10 dB) & 2.02 & 4.25
  & 10.34 & \bf 4.81 & 3.05 
  & \bf 10.92 & 4.42 & 2.76 
  & \bf 12.78 & \bf 4.42 & 2.68 
\\
\bottomrule
\end{tabular}}
\end{center}
\vspace{-3mm}
\label{tab:main_results}
\end{table*}
\begin{table}[!t]
\small
\caption{WER (\%) of effect of augmentation in self-distillation to make noisy pair. Fine-tuning was performed on clean data.}
\vspace{-3mm}
\renewcommand{\tabcolsep}{5pt}
\def\arraystretch{1.0}
\begin{center}

\scalebox{0.9}{\begin{tabular}{lccccc}
\toprule
\multirow{2}{*}{\textbf{Method}} & \multicolumn{3}{c}{\textbf{Augmentation}} & \multirow{2}{*}{\textit{test-clean}} & \multirow{2}{*}{\textit{test-other}}\\[-2pt]
\cmidrule(lr){2-4}
&SpecAug & Noise & RIR & \\
\midrule
Baseline &&& & 2.58 & 5.41 \\
Fine-tuning &\multicolumn{3}{c}{} & 2.00 & 4.45 \\
\midrule
\multirow{5}{*}{DASH} & \checkmark & & & 2.01 & 4.16 \\
&\checkmark & \checkmark & & {1.99} & \textbf{4.10} \\
&\checkmark & & \checkmark & 1.99 & 4.12 \\
&&\checkmark & \checkmark & \textbf{1.97} & 4.18 \\
&\checkmark & \checkmark & \checkmark & 2.01 & 4.18 \\
\bottomrule
\end{tabular}}
\end{center}
\vspace{-2mm}
\label{tab:aug_effect}
\end{table}

\section{Experimental Setup}
%
%
\subsection{Dataset and Augmentation}
\label{sec:dataset}

For training and dataset, we use LibriSpeech~\cite{panayotov_librispeech_2015} \textit{train-960} for supervised ASR training and LibriLight Medium~\cite{kahn_libri-light_2020} for encoder-only self-distillation. For evaluation, we used greedy decoding without external language models and report WER on \textit{test-clean} and \textit{test-other} of LibriSpeech.

During self-distillation, the clean view was left unaugmented while the noisy view is constructed using augmentations. We consider three strategies: (i) SpecAugment~\cite{park_specaugment_2019}, (ii) additive noise mixing with segments from MUSAN~\cite{snyder_musan_2015} at signal-to-noise ratios (SNRs) uniformly sampled between -5 and 15~dB, and (iii) reverberation by convolution with mono room impulse responses (RIRs) from the DNS Challenge 2021 dataset~\cite{reddy_interspeech_2021}.

To select the default noisy-view construction, we compare several augmentation combinations in Table~\ref{tab:aug_effect}. While DASH performs robustly across all choices (yielding consistent gains on \textit{test-other}), SpecAugment+Noise provides the strongest overall trade-off and achieves the best \textit{test-other} result among the tested variants. We therefore adopted SpecAugment+Noise as our default setting throughout the experiments. We also observe that using all three augmentations simultaneously slightly degraded performance, suggesting that overly strong corruption can blur phonetic cues.

%
%
\subsection{Implementation Details}
\label{sec:implementation-details}

We adopt nvidia/parakeet-tdt\_ctc-110m as our baseline model, a hybrid ASR architecture combining Token-and-Duration Transducer (TDT)~\cite{xu_efficient_2023} and CTC~\cite{graves_connectionist_2006}. 
The encoder consists of 17 FastConformer~\cite{rekesh_fast_2023} layers with 512-dimensional hidden representations. 
The model uses a hybrid loss function combining TDT and CTC as $\mathcal{L}_{\text{ASR}} = 0.7 \cdot \mathcal{L}_{\text{TDT}} + 0.3 \cdot \mathcal{L}_{\text{CTC}}$. 
We applied a dropout rate of 0.1 for regularization.
All architectural configurations followed the default settings provided by the Parakeet toolkit.

In self-distillation stage, the encoder from the model was trained with our proposed loss $\mathcal{L}_{\text{DASH}}$. We applied early stopping after 5k steps, corresponding to $\sim$3{,}230 hours of unlabeled speech from the LibriLight Medium. 
We froze the prediction and joint networks of the baseline model and update only the encoder. 
For fine-tuning, the full model was trained for 100k steps on LibriSpeech \textit{train-960} with the ASR loss $\mathcal{L}_{\text{ASR}}$ using the same optimizer configuration and batch settings. All experiments were conducted on two NVIDIA GeForce RTX 3090 GPUs. 
Notably, the additional cost of the self-distillation stage was marginal: on a single RTX~3090, fine-tuning takes roughly 12 hours, whereas self-distillation takes about half an hour (i.e., $\sim$4\% of the fine-tuning time).

For prototype construction, we extracted encoder representations from 100,000 randomly sampled utterances from the LibriLight Medium and clustered them using k-means with $K\hspace{-.5mm}=\hspace{-.5mm}512$. We used temperature-scaled KL divergence ($\tau_{temp}\hspace{-.6mm}=\hspace{-.6mm}3.5$) to match student-teacher distributions. The teacher model was updated via EMA with a decay rate of $\alpha=0.999$. Training used the AdamW optimizer with a learning rate $5 \times 10^{-5}$, weight decay $10^{-4}$, and $\beta$ parameters [0.9, 0.999]. We apply gradient clipping with a L1-norm of 1 and employed Lhotse dynamic bucketing with a batch duration up to 500 seconds. 

\section{Experimental Results}
%
%

\subsection{Results on SNR Range}

Table~\ref{tab:main_results} presents WER evaluations of the baseline, the standard fine-tuning approaches, and the proposed DASH framework across \textit{test-clean}, \textit{test-other}, and simulated noisy conditions. NOISEX-92 \cite{varga_assessment_1993} was used for noise mixing, which consists of babble, pink, and white noise. 
Consistent with our preliminary hypothesis, simply incorporating noise during standard supervised fine-tuning introduced a trade-off. While the fine-tuning only model trained on noisy data (0 to 15 dB) improved noise robustness compared to clean fine-tuning, it simultaneously degraded the performance on \textit{test-clean}. Conversely, DASH mitigated this trade-off. The DASH model (pre-trained and fine-tuned on 0 to 15 dB noise) not only outperformed its fine-tuning counterpart across all noisy conditions but also achieved the best overall performance on \textit{test-clean}. 
Across all configurations, DASH yielded competitive \textit{test-clean} results, closely matching or even surpassing the clean-only fine-tuning baseline.
This demonstrates that our prototype-based self-distillation effectively extracts clean speech characteristics, preventing the model from overfitting to specific noise patterns and thereby preserving clean baseline accuracy. 

The inherent advantage of our decoupled two-stage paradigm is exhibited in the DASH (Noisy $\rightarrow$ Clean) configuration. Although the model had been unexposed to noisy data during the supervised fine-tuning phase, it exhibited substantial noise robustness compared to all the fine-tuning-only models. This confirms that the label-free pre-training phase successfully establishes noise-invariant representations independently of the ASR objective. 
Furthermore, DASH demonstrates better generalization to unseen acoustic variations and challenging speaker characteristics, as shown in \textit{test-other} results. While standard fine-tuning approaches struggled to drop below 4.35\%, all DASH configurations consistently achieve significant improvements, yielding WERs between 4.10\% and 4.25\%. Because the pre-training phase comprehensively handles the robust feature extraction, the model maintained high stability and generalization on out-of-domain data (\textit{test-other}), independent of the specific augmentation strategies employed during the fine-tuning.

%
%

\subsection{Effect of EMA and Layer Selection}

\begin{table}[!t]
\vspace{-2mm}
\caption{Ablation on teacher update and layer selection in self-distillation, varying EMA update scheme (step-wise vs.\ 1000-step vs.\ frozen) and the distillation depth (multi-layer vs.\ final layer only).}
\vspace{-2mm}
  \label{tab:ema_update}
  \centering
  \small
  \renewcommand{\arraystretch}{1.05}
    \scalebox{0.9}{\begin{tabular}{lccc}
    \toprule
    \textbf{Method} 
    & \textit{test-clean}
    & \textit{test-other} \\
    \midrule
    Baseline & 2.58 & 5.41 \\
    Fine-tuning only & 2.14 & 4.40 \\
    \midrule
    \rowcolor{Gray}
    {DASH (Self-Distillation)}  & \textbf{2.02} & \textbf{4.25} \\
    \quad w/ EMA (1000-step interval) & 2.04 & 4.27 \\
    \quad w/o EMA (freeze) & {2.05} & {4.30} \\
    \quad w/o EMA (freeze) \& Single-layer(17) & {2.10} & {4.45} \\
    \bottomrule
  \end{tabular}}
\vspace{-2mm}
\end{table}

Table~\ref{tab:ema_update} highlights the importance of the EMA update strategy and the multi-layer architecture during the self-distillation phase. The analysis reveals three critical insights regarding the teacher-student dynamic.
The proposed DASH framework, which continuously updates the clean (teacher) network at every single step using EMA, yielded the most robust and optimal performance across all test sets. This confirms that a smoothly and continuously evolving teacher provides the most stable and effective guidance for the student network.
Also, reducing the frequency of the EMA updates or freezing the teacher network entirely results in a consistent degradation of recognition accuracy. Without the continuous feedback loop provided by the step-wise EMA, the static or delayed teacher failed to adapt its target representations, thereby limiting the student model's ability to learn generalized acoustic features.

The most severe performance drop occurred when the teacher network was frozen and the distillation was restricted solely to the final encoder layer. Under this constrained setting, the model's generalization capability on challenging acoustic conditions significantly deteriorated, even falling behind the standard fine-tuning baseline. These results indicate that our multi-layer distillation approach is indispensable; distilling intermediate hierarchical representations is just as critical as maintaining a dynamic teacher for acquiring noise-invariant speech characteristics.

\begin{figure}
    \centering
\vspace{-2mm}
\subfloat[fine-tuning only (Layer 6)]{\includegraphics[width=0.23\textwidth]{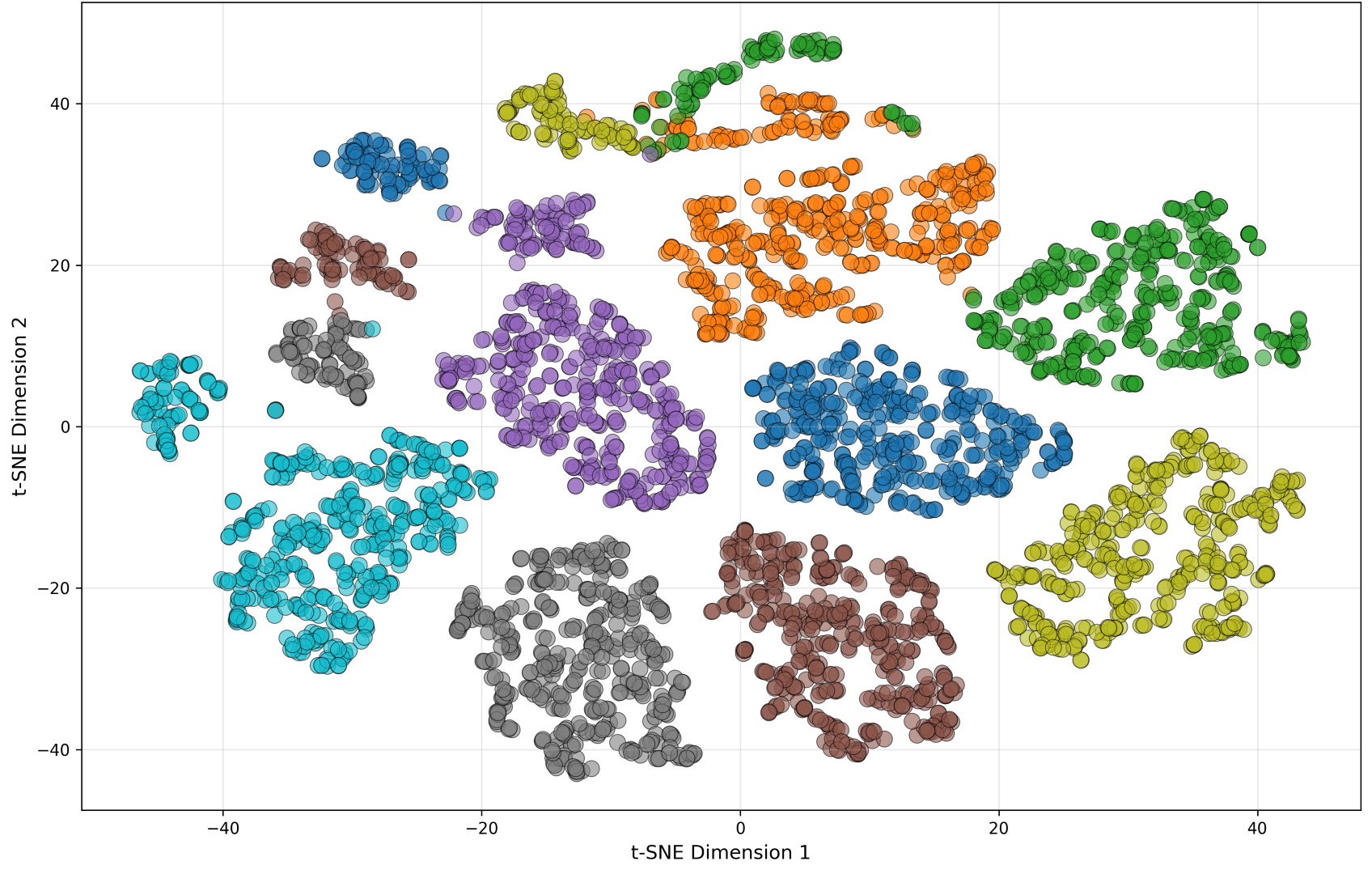}}\label{fig:sp0}\hspace{0mm}
\subfloat[DASH (Layer 6)]{\includegraphics[width=0.23\textwidth]{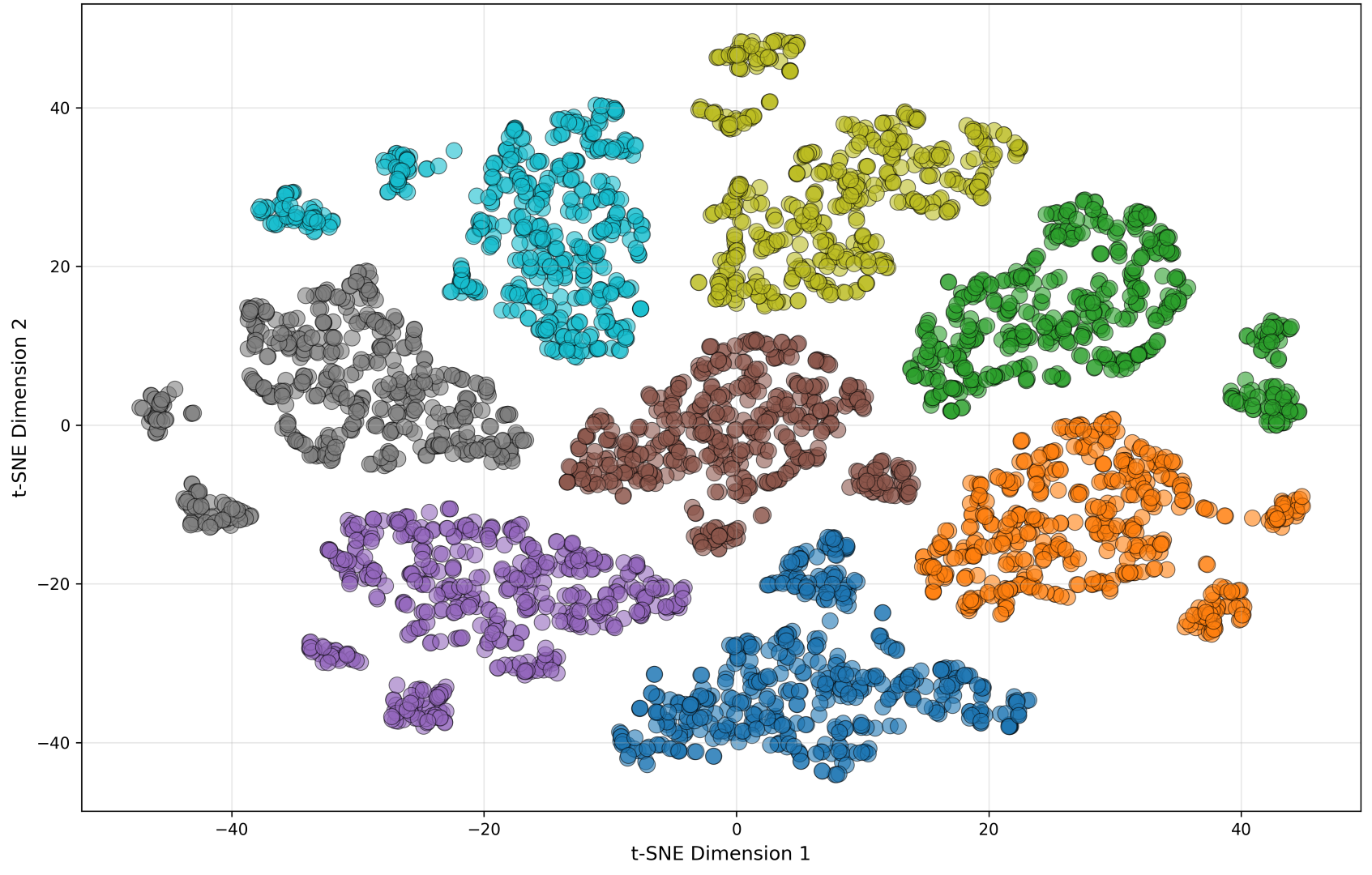}}\label{fig:sp1}\hspace{0mm}
\subfloat[fine-tuning only (Layer 17)]{\includegraphics[width=0.23\textwidth]{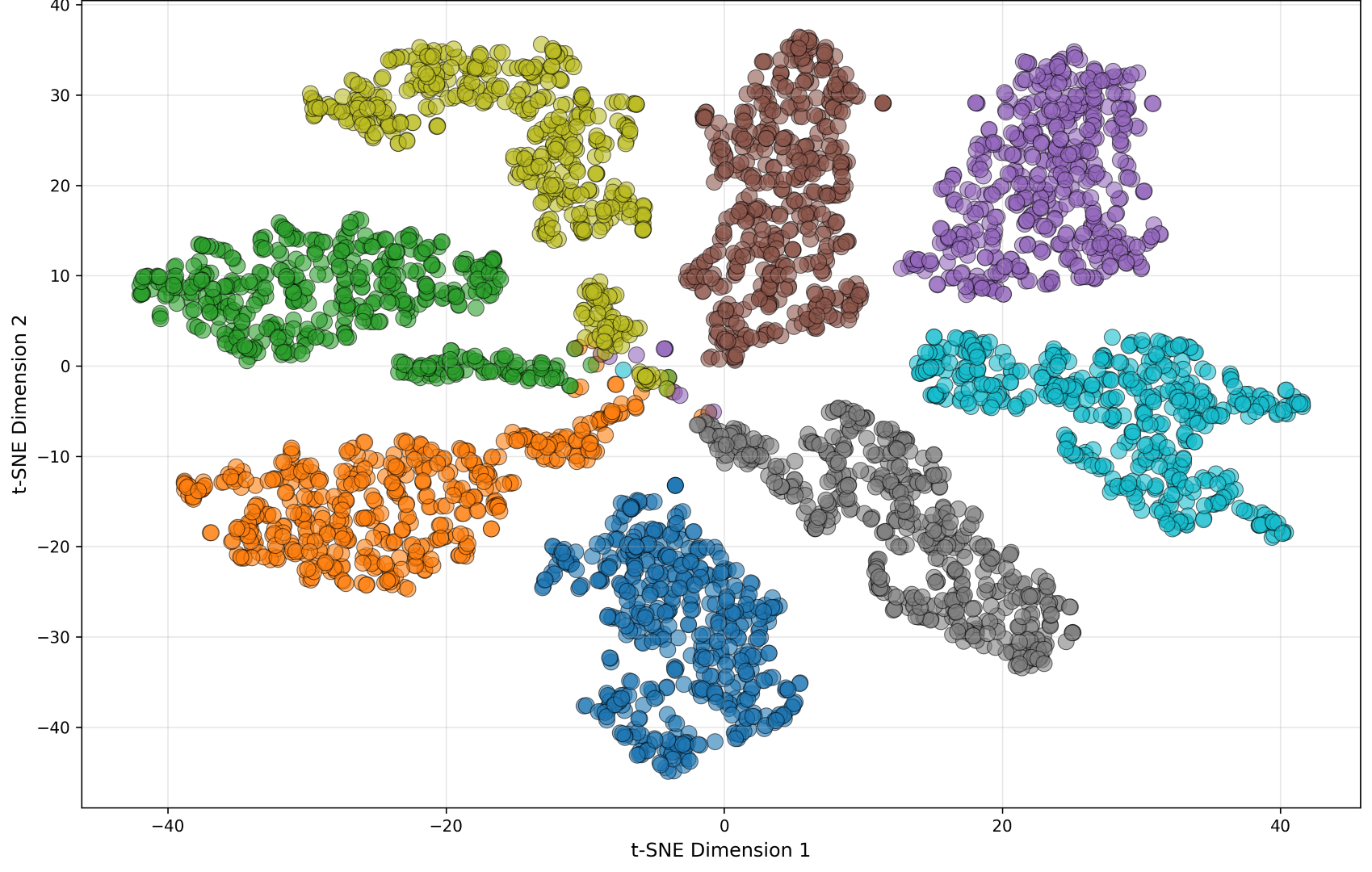}}\label{fig:sp2}\hspace{0mm}
\subfloat[DASH (Layer 17)]{\includegraphics[width=0.23\textwidth]{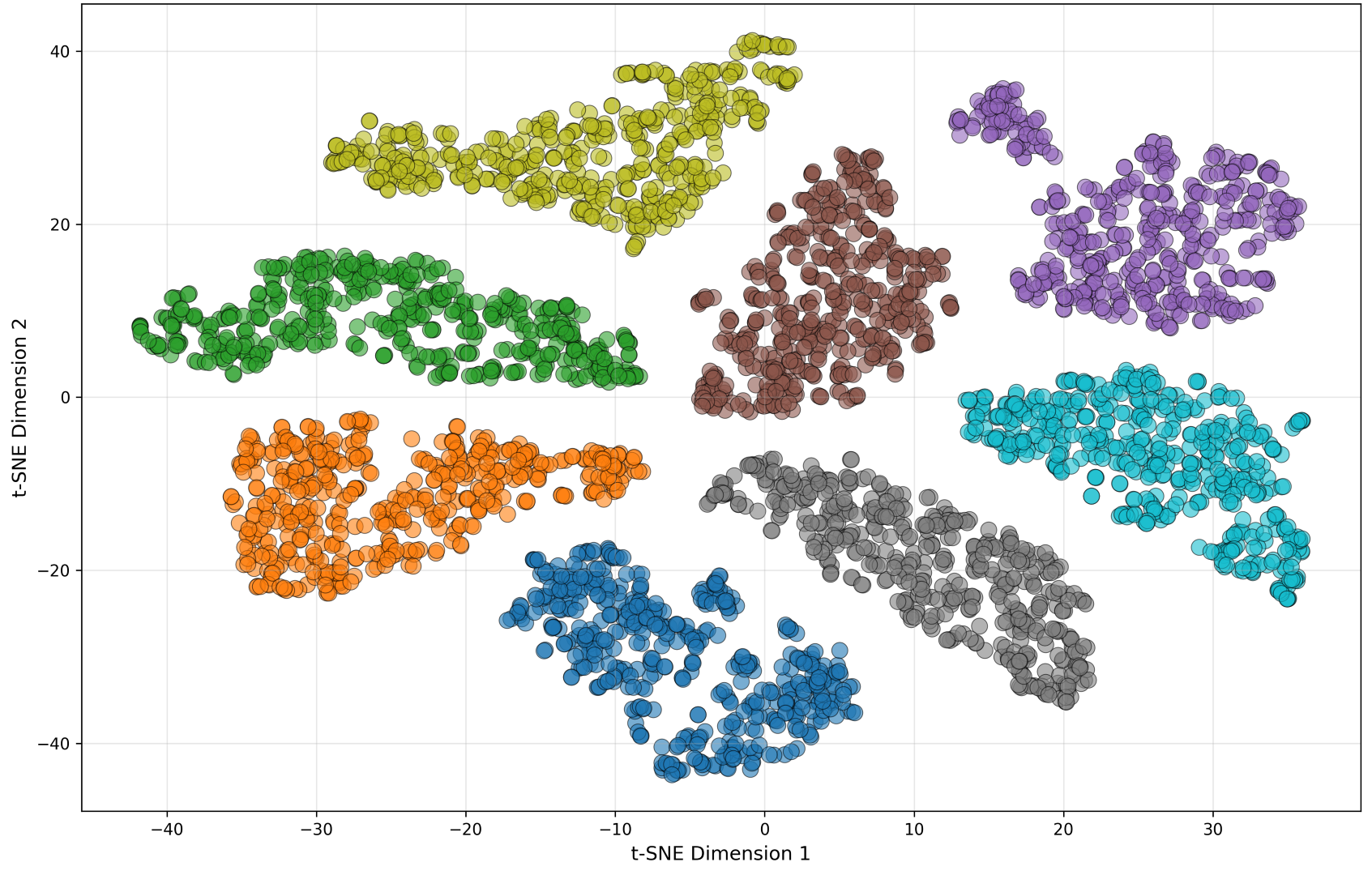}}\label{fig:sp3}\hspace{0mm}

\vspace{-2mm}
\caption{t-SNE visualization of encoder representations extracted from Layers 6 and 17 for the fine-tuning-only baseline (left) and DASH (right). Points are colored by utterance.}
\label{fig:tsne}
\vspace{-3mm}
\end{figure}

\subsection{Visualization of Noise-Invariance}

To qualitatively assess noise-invariance, we visualized encoder representations using t-SNE. We randomly selected several utterances and added noise from the NOISEX-92 with SNR of 0 dB, then extracted representations from different encoder layers (Layers 6 and 17) and projected them onto 2D. As shown in Figure~\ref{fig:tsne}, the fine-tuning-only baseline exhibits broader and sometimes fragmented clusters. Specifically, in Layer 6, while representations from both models show some susceptibility to noise, DASH groups these low-level acoustic features much more cohesively. In Layer 17, although both models naturally form more refined semantic boundaries, DASH still achieves noticeably tighter clusters, effectively eliminating the residual noise entanglement seen in the baseline. Overall, this consistent layer-wise behavior indicates that DASH robustly aligns noisy examples with their clean counterparts across both acoustic and semantic dimensions.


\section{Conclusion}

In this paper, we proposed DASH, a dual-view self-distillation framework that improves ASR robustness by enforcing clean--noisy consistency. DASH adopts a decoupled two-stage pipeline: label-free encoder pre-training with an EMA teacher and prototype-based KL distillation across multiple encoder layers, followed by supervised fine-tuning. Experiments on LibriSpeech with noisy mixtures demonstrate that DASH consistently reduces Word Error Rate (WER) under diverse noise types and SNRs while preserving clean baseline accuracy, effectively mitigating the robustness--clean trade-off typically observed in noise-augmented fine-tuning. Furthermore, our ablation studies verify that step-wise EMA updates and multi-layer distillation are crucial components for achieving stable performance gains.

\pagebreak

\section{Acknowledgments}
This work was partly supported by Institute of Information \& communications Technology Planning \& Evaluation(IITP) grant funded by the Korea government(MSIT)(RS-2022-II220989, Development of Artificial Intelligence Technology for Multi-speaker Dialog Modeling) and National Research Foundation of Korea(NRF) grant funded by the Korea government(MSIT)(RS-2026-25470024, Research on Spatial Audio Reasoning with Large Language Models across Diverse Microphone Arrays). 

\section{Generative AI Use Disclosure}
Generative AI tools were used solely for editing and polishing the manuscript to improve linguistic clarity. These tools were not employed to produce any significant portion of the technical or scientific content, and the authors remain fully responsible and accountable for the integrity and final results of the work.

\bibliographystyle{IEEEtran}
\bibliography{mybib}

\end{document}